# Near resonant nanosecond laser driven nonlinear optical response in As$_{50}$S$_{50}$ thin films

Dipendranath Mandal[1†], Pritam Khan[2†] and K. V. Adarsh[1*]

[1]Department of Physics, Indian Institute of Science Education and Research, Bhopal 462066, India

[2] Department of Physics and Bernal Institute, University of Limerick, V94 T9PX Limerick, Ireland

**ABSTRACT:** Nanosecond near resonant excitation in As$_{50}$S$_{50}$ thin films leads to strong nonlinear optical response, i.e. nonlinear absorption coefficient up to $4 \times 10^6$ cm/GW and nonlinear refractive index of 8.5 cm$^2$/GW, both of which is the strongest ever reported in amorphous semiconductors. We propose a five-level energy model to explain such effect which indicates that nonlinear process is reverse saturable absorption in nature, mediated by excited state absorption from triplet-triplet transition. On the other hand, observation of negative nonlinear refractive index reveals the occurrence of self-defocusing effect. Finally, benefitting from the strong nonlinear response, we demonstrate a promising application of As$_{50}$S$_{50}$ thin films as an optical limiter for optoelectronic sensors.

† Authors contributed equally.

* Author to whom correspondence should be addressed. Electronic mail: adarsh@iiserb.ac.in.

# 1. Introduction

In recent years, there is a huge surge in understanding the third order nonlinearities of optical materials from both fundamental and plethora of application aspects, including, optical switching [1], optical modulation [2], optical limiting [3, 4] etc. Among these, optical limiter is of particular importance because it helps in protecting human eye as well as optical sensors in optoelectronic application from intense laser source [4]. In principle, an ideal optical limiter must have high transmission in low intensity beam while reduced transmission against intense beam, over a wide dynamic working range. Optical limiting is typically associated with nonlinear optical process, e.g. multiphoton or excited state absorption (ESA) [5-8]. Therefore, the quest for materials with high optical nonlinearity, i.e. high nonlinear absorption ($\beta$) and high nonlinear refractive index ($n_2$) remains a central research question over many years. A wide range of amorphous and crystalline materials, for example, porphyrin [9], ZnO nanorods [10], plasmonic materials [11], nanocomposites of transition metal dichalcogenides [12], graphene oxide [4, 13] etc. have shown their applicability to some extent in optical limiting. However, fabricating such structures is a real challenge because it involves complicated synthetic routes that requires utmost precision. In this regard, amorphous chalcogenide glasses (ChGs) and their thin films hold advantages because of their easy synthetic procedure via thermal evaporation technique [14, 15]. In addition to that, ChGs possess high linear refractive index, high third order nonlinear refractive index ($n_2$) and ultrafast response time [16]. ChGs are formed by chalcogen elements (S, Se, Te) in combination with elements like As, Ge, Ga, Sb etc [14, 15, 17]. Among the family of ChGs, As-based compositions are preferable owing to their high chemical stability. Compared to the conventional $As_{40}S_{60}$ ($As_2S_3$), which has mean coordination number (MCN) 2.4, in our present study we choose over-constrained $As_{50}S_{50}$ ($As_2S_2$), with MCN 2.5, because of their intrinsic structural rigidity and thermal stability against pulsed laser.

In this article, we employed conventional Z-scan technique to demonstrate strong nonlinear optical response in $As_{50}S_{50}$ thin films. A comparative study reveals that both $\beta$ and $n_2$, are the highest ever reported in the family of amorphous semiconductors. ESA observed in our sample from triplet-triplet transition is explained from a five-level energy model. We also showed that the strong nonlinear effects in $As_{50}S_{50}$ hold potential applications in fabricating optical limiting devices.

## 2. Methods

### 2.1 Sample preparation

$As_{50}S_{50}$ thin films were prepared by using conventional melt-quenching method starting from 99.999 % pure As and S. The cast samples were used as the source material for depositing thin films on microscopic glass substrate by thermal evaporation technique. To preserve the target stoichiometry from the starting bulk material, we used a low deposition rate of 2-5 A°/s. We have later performed EDAX and confirmed the uniformity of composition is within the experimental error of this technique, i.e., within 2–3%. We prepared film of thickness ~ 420 nm which was verified by using ellipsometric measurements [18]. X-ray diffraction (XRD) measurements further indicates that the sample is amorphous in nature.

### 2.2 Raman spectroscopy

The Raman spectra are obtained in a Horiba JY LabRam HR Evolution Spectrometer mounted with a grating of 1800 grooves/mm using a 50 X objective with N.A. = 0.5 lens in back-scattering geometry. We excite the sample by a diode laser of wavelength 532 nm with spot diameter of ~ 0.5 μm. The detection is performed by an air-cooled charge coupled device (CCD) detector. To avoid local heating by the laser, measurements are performed at very nominal power of ~ 1 mW.

## 2.3 Z-scan measurements

To unveil the third order nonlinear optical response of As$_{50}$S$_{50}$ thin films, we employed Z-scan technique which is a precise method based on self-focusing. A schematic diagram of the Z-scan set up is shown in Fig. 1. In a typical Z-scan set up transmission is measured in the far field as a function of sample position Z, along the laser propagation of a focused laser beam. In our present study, we measured transmission in two different configurations. In the first case, known as "open-aperture" Z-scan, all the transmitted laser beam is collected and measured after the sample. In the second case, laser light is transmitted through small aperture placed in the far field, which is defined as "closed-aperture" Z-scan. By performing open and closed aperture Z-scan measurements, we can determine nonlinear absorption coefficient and non-linear refractive index of the films, respectively. We excite the sample with the second harmonics (532 nm) of Nd:YAG laser of pulse width 7 ns. To exclude the possibility of heating and photo damage, we used a fixed repetition rate of 10 Hz [19]. We expect that for 10 Hz repetition rate, the temporal separation between two pulses in 100 ms which is reasonably long to dissipate the heat and avoid progressive heating before the arrival of next pulse to the sample. The beam is focused on the axis of a lens of focal length 20 cm where the film is moving by computer controlled translational stage. The Rayleigh length and beam waist are measured to be 3.7 mm and 25 μm, respectively. The linear transmission of the film on glass is ≈ 60% measured in the far field region.

## 2.4 Nanosecond Pump-probe measurements

For pump probe transient absorption (TA) measurements, we used second harmonics of the Nd:YAG laser (7-ns pulses centered at 532 nm with an average fluence of 26 mJ/cm$^2$ and having a repetition rate of 10 Hz) as pump in single-shot mode. The probe beam was selected from a Xenon Arc lamp (120 W) using a holographic grating with 1200 grooves/mm and delayed with respect to the pump beam using a digital delay generator. The pump and probe beams were overlapped at the sample. The change in absorbance of the probe beam, $(\Delta A = -\log[I_{es}/I_{gs}])$ at different delays was recorded using a photomultiplier tube and a digital oscilloscope. Here $I_{es}$ and $I_{gs}$ are the transmitted intensities of probe beam after delay time t following the pump beam excitation and in ground state, respectively.

## 3. Results and Discussions

To analyze the structural composition of our $As_{50}S_{50}$ thin films, first, we recorded Raman spectra as shown in Fig. 2. Clearly, we observe two intense broad peaks centered at 348 and 235 cm$^{-1}$ followed by two residual peaks at 275 and 187 cm$^{-1}$. The Raman band centered at 348 cm$^{-1}$ is assigned to the vibrations of $AsS_3$ pyramids with minor contributions from $\alpha$-$As_4S_4$, $\beta$-$As_4S_4$, $\chi$-$As_4S_4$, $\alpha$-$As_4S_3$, and $\beta$-$As_4S_3$ [20]. From this, it is clear that the assignment of Raman modes to individual structural units is quite complicated due to high number of possible Raman active vibrational modes and their strong overlapping. Another strong Raman band around 235 cm$^{-1}$ is attributed to the As-As vibrations in $As_4S_4$ molecules. The weak peak at 275 cm$^{-1}$ is assigned to pararealgar ($\gamma$-$As_4S_4$) and $\chi$-$As_4S_4$. The other weak peak at 187 cm$^{-1}$ belongs to $\beta$ -$As_4S_4$.

To select the excitation wavelength for performing Z-scan measurements, first we measured the bandgap of the sample. In this context, Fig. 3(a) shows the optical absorption spectra of the film. Since $As_{50}S_{50}$ is an indirect bandgap material we use the following Tauc equation to calculate the bandgap of the sample as:

$$(\alpha h \nu) = B(h \nu - Eg)^2 \tag{1}$$

where α, h, ν, $E_g$ and B are the absorption coefficient, Plank′s constant, frequency, optical band gap and a band tailing (Tauc) parameter respectively. The intercept of the straight line at the photon energy axis will provide $E_g$ as shown in Fig. 3(b). Best fit of the experimental data indicates that the bandgap of the sample is $2.05 \pm 0.01$ eV.

Next, we performed both open and closed-aperture Z-scan measurements with near bandgap 532 nm excitation laser. In this regard, Fig. 4(a) shows the open-aperture Z-scan traces of $As_{50}S_{50}$ thin films at peak intensities of 10 and 30 MW/cm$^2$, which are measured at the focal point (z=0). To obtain reference signal, first we recorded the Z-scan measurements of bare glass substrate without any coating of $As_{50}S_{50}$ thin films, represented by solid blue circles in Fig. 4(a). For bare substrates, we kept the intensity at 50 MW/cm$^2$ which is higher than the intensity used for our nonlinear measurements on thin films to discard any contribution from the substrate. Clearly a horizontal line is observed before and after the focal point which indicates that the bare substrate does not contribute to the nonlinear optical response of $As_{50}S_{50}$. For $As_{50}S_{50}$ thin films, it can be seen from the figure that at both the intensities, normalized transmittance exhibits gradual decrease while the sample moves towards focal point, and reaches minimum at z = 0. Such nonlinear response indicates that $As_{50}S_{50}$ undergoes reverse saturable absorption (RSA) similar to the observation by Elim *et al* [21]. It is important to note that, Z-scan curves does not exhibit any asymmetry before and after the focal point, i.e. transmission value remains the same, which indicates that the sample doesn't undergo any thermal or photo damage upon nanosecond laser illumination.

In semiconductors, RSA can be classified into two types; first is two-photon absorption (TPA) and second is excited state absorption (ESA) [22]. Off late, Poornesh and co-workers explain the RSA observed in organic molecule by assuming a five-level model [23]. However, in solid state

As$_{50}$S$_{50}$ thin films, the RSA can be explained from the energy level diagram as shown in Fig. 4(b). In general, non-linear optical response can be explained in terms of two physical process, intraband (within the band) and interband (between two bands) transitions. Let's assume, time constant related to intra and interband transitions are $\tau_1$ and $\tau_2$. In the present case, bandgap (E$_g$) of As$_{50}$S$_{50}$ thin film is found to be 2.05 eV which is less than the excitation photon energy 2.33 eV (532 nm). Therefore, both intra and interband transitions can contribute to the overall RSA observed in As$_{50}$S$_{50}$. However, as we used nanosecond laser pulses to excite the sample, pulse width (7 ns) is much longer than the intraband relaxation time $\tau_1$, which is approximately of the order of femtoseconds to sub-picoseconds time frame. So, we can discard the possibility of intraband transition in our sample. Likewise, the only plausible mechanism remains interband transition between valence band (ground sate) and conduction band (excited state). In general, when the values of $\tau_2$ are smaller, the excited carriers can quickly revert back to the ground state, consequently strong ground-state absorption gives rise to saturable absorption (SA). On the other hand, for larger $\tau_2$ values, i.e. when the excited carriers slowly return back to the ground state, excited state absorption dominates and the sample exhibits RSA. As$_{50}$S$_{50}$, being a ChG possess many deep and shallow trap states, between the bands. In such cases, excited carriers may not come back to ground state directly, rather they trapped into a metastable state via non-radiative decay. To find the interband relaxation time, we performed the nanosecond transient absorption (TA) measurements in As$_{50}$S$_{50}$ thin films. In this regard, temporal evolution of TA at selected wavelengths are shown in Fig. 4(c). Quite clearly, for some wavelengths, TA is reversible, indicating that excited carriers come to back to the ground state whereas for some other TA is irreversible which indicates that the carriers get trapped in the metastable state. In this regard, Table 1 reveals the interband relaxation time for the selected wavelengths following TA measurements. From the $\tau_2$ values we found that excited carriers non-radiatively decay to the metastable trapped state or ground state within few microseconds which is much longer than the pulse duration of ns laser. The obtained results are similar to our recent work in As-based ChG films, i.e. $As_{35}Se_{65}$ [24] and $As_{40}S_{60}$ [17] and $Ge_5As_{30}Se_{65}$ [ 25] thin films. Consequently, excited state cross section remains higher than ground state cross section and the sample exhibits RSA.

In this context, we calculated the absorption cross sections for excited and ground state from the following equations:

$$\sigma_{es} = -\frac{\log T_{max}}{NL} \qquad (2)$$

$$\sigma_{gs} = -\frac{\log T_0}{NL} \qquad (3)$$

here $T_0, T_{max}, N, L$ are the linear transmittance of the material, transmittance peak of the saturable absorption, carrier density of the material at the ground state and the thickness of the sample, respectively. In $As_{50}S_{50}$, calculated values of $\sigma_{es}$ and $\sigma_{gs}$ are found to be $2.25 \times 10^{-12}$ cm$^2$ and $1.28 \times 10^{-12}$ cm$^2$, respectively. Consequently, we found that the ratio $\sigma_{es}/\sigma_{gs}$ (= 1.76) is greater than unity, which indicates that the nonlinear optical process is RSA from excited state absorption (ESA).

To quantify the observed ESA, we used the following propagation equation in the dispersion as a function of the position as:

$$\frac{dI}{dz} = -\alpha(I)I \qquad (4)$$

where I and z are the intensity of the laser beam and propagation distance inside the sample. $\alpha(I)$ is the intensity dependent absorption coefficient which is defined by:

$$\alpha(I) = \frac{\alpha_0}{1+\frac{I}{I_s}} + \beta_{ESA}I \qquad (5)$$

where $\alpha_0$, $I_s$ and $\beta_{ESA}$ are the linear absorption coefficient, saturable intensity and the ESA coefficient, respectively. To obtain more quantitative picture, we exploit Z-scan theory by expressing normalized transmittance as a function of position z as:

$$T_N = \frac{1}{q_0\sqrt{\pi}} \int_{-\infty}^{+\infty} \ln\left(1 + q_0 e^{-t^2}\right) dt \qquad (6)$$

Where $q_0 = \frac{\beta I_0 L_{eff}}{1+\frac{Z^2}{Z_0^2}}$ and $L_{eff} = \frac{(1-e^{-\alpha L})}{\alpha}$. The best fit to the normalized transmittance provides us $\beta_{ESA}$. In this regard, Fig. 4(d) shows the variation of $\beta_{ESA}$ with input peak intensity ($I_0$). If the non-linear process is associated with simple TPA alone, $\beta_{ESA}$ should not exhibit any variation with $I_0$. But in a start contrast, $\beta_{ESA}$ decreases with increase in $I_0$ which is attributed to sequential TPA via ESA (RSA) similar to the observation of Couris [26] and Bindhu [27]. Our results indicate that $\beta_{ESA}$ is found to be $(4.3 \pm 0.7) \times 10^6$ cm/GW at the peak intensity of 10 MW/cm². In this regard, all nonlinear parameters are tabulated in the Table 2. Such enhanced ESA in our sample intrigues us to compare the present result with the previous reports and the results are summarized in Table 3 [28-32]. Quite clearly, the amplitude of $\beta_{ESA}$ is ~ 6 order ($10^6$) higher than previously obtained results and also surpass the nonlinearity obtained in our last measurements on As-Sb-Se system by a factor of 10 [19]. Such extremely high value of non-linear absorption coefficient of As₅₀S₅₀ indicates its potential use in the field of data transfer through waveguide without much loss and an ideal candidate for designing optical limiting devices.

Observation of remarkable ESA with very high $\beta_{ESA}$ motivates us to explore the potential of our sample in device fabrication, specifically in optical limiting process. An optical limiter is a selective nonlinear device which attenuates intense laser beam while allowing low intensity beams. They also play a crucial role in protecting optoelectronic devices, e.g. photomultiplier tube, photodiode and human eye from intense laser. To demonstrate the device potential, we have plotted in Fig. 5(a) the variation of transmitted output intensity ($I_{out}$) as a function of input intensity ($I_{in}$). Quite clearly, at lower intensities, $I_{out}$ scales a linear relationship with $I_{in}$, following Beer Lambert law, thus the device remains inactive. However, as $I_{in}$ increases, $I_{out}$ exhibits deviation from linearity as shown by the dashed line in Fig. 5(a) which shows that $As_{50}S_{50}$ thin film limits $I_{out}$ for all higher $I_{in}$ above the certain threshold of 5 MW/cm$^2$. This particular value of $I_{in}$ is very important to determine the performance of an optical limiter. As the experimental results are based on the Z-scan set up, the reliability of the optical limiter is also checked by plotting the normalized transmittance as a function of the input intensity as shown in Fig. 5(b). It can be seen that the transmission starts decreasing when input laser intensity ($F_{on}$) exceeds 5 MW/cm$^2$, as in Fig. 5(a) which is a necessary criterion for the working principle of an optical limiter.

After demonstrating large non-linear absorption coefficient by open-aperture Z-scan measurements, we performed closed-aperture Z-scans to determine the sign and magnitude of the nonlinear refractive index $n_2$ of $As_{50}S_{50}$ thin films. We adjusted the size of the aperture in front of the detector in such a way that the transmission reduced to one third of the incident value. Typical normalised closed aperture Z-Scan traces is shown in Fig. 6 at peak intensity of 30 MW/cm$^2$. Our sample exhibits pre-focal transmittance maximum (peak) followed by a post-focal transmittance minimum (valley), i.e. peak-valley (P-V) behaviour which is associated with self-defocussing effect and characterized by negative $n_2$. The nonlinear refractive index $n_2$ is calculated using the equation:

$$n_2 = \frac{\lambda}{2\pi} \frac{\Delta\phi_0}{I_0 L_{eff}} \qquad\qquad (7)$$

where $I_0$ is the peak intensity at the focus and $L_{eff}$ is the effective thickness of the sample, expressed as $L_{eff} = \frac{(1-e^{-\alpha L})}{\alpha}$, where $L$ is the thickness of the sample and $\alpha$ is the absorption coefficient. For a transparent sample, we assume $L_{eff} \approx L$. $\Delta\phi_0$ is the on-axis phase change, determined by fitting the experimental data with the equation:

$$|\Delta\phi_0| = \frac{\Delta T_{p-v}}{0.406\,(1-s)^{0.25}} \qquad (8)$$

where $\Delta T_{p-v}$ is the difference in magnitude between the normalized transmittance at peak ($T_p$) and at the valley ($T_v$). S is the transmittance of the aperture without any sample and is represented as $S = 1 - \exp\left(-\frac{2r_a^2}{\omega_a^2}\right)$, where $r_a$ is the aperture radius and $w_a$ is the beam radius at the aperture. From our calculation the axial peak valley difference is 2.65 $Z_R$, which is greater than 1.7 $Z_R$, which gives a clear indication that the observed nonlinear process has thermal contribution and can be assigned as a third order process ~~with~~ as proposed by Nagaraja and co-workers [22]. In order to extract the contribution purely from nonlinear refraction we used the division method described by Sheik-Bahae *et. al* [33] and fitted the experimental data in Fig. 6 with the equation:

$$T(x) = 1 + \frac{4x\Delta\phi_0}{(1+x^2)(9+x^2)} + \frac{4(3x^2-5)\,\Delta\phi_0^2}{(1+x^2)(9+x^2)(25+x^2)} \qquad (9)$$

where T(x) is the transmittance of the sample and $x = \frac{z}{z_r}$ is the fraction of axial distance with Rayleigh length. Our calculation shows that $n_2$ is ~ (8.49 ± 0.81) cm$^2$/GW which is 7 orders of higher in magnitude as compared to the reported value of fused silica $(3.0 \pm 0.4) \times 10^{-7}$ cm$^2$/GW. The ChGs exhibit strong nonlinear response than other counterparts because of their low vibrational bond energy induced by heavy chalcogen atoms. Consequently, ChGs become optically transparent up to mid IR wavelength range [34]. Apart from that glass density of the ChGs are also higher which when combined with strong polarizability, gives rise to high linear refractive index n ∼ 2–3 [34, 35]. Likewise, higher linear refractive index following empirical Miller's rule [36] exhibit high nonlinear refractive index, as well as strong nonlinear optical response [35].

In summary, we demonstrated that nonlinear response in $As_{50}S_{50}$ thin films can be successfully exploited in fabricating the optical limiting device. Intensity dependent open aperture Z-scan curves indicates that non-linear process is consistent with RSA. A five-level energy model diagram explain that RSA is originated from the ESA via triplet-triplet transition. On the other hand, closed aperture Z-scan measurements indicates that $As_{50}S_{50}$ thin films has negative $n_2$ which is associated with self-defocussing effect. Finally, experimental results obtained in terms of β and $n_2$ are found to be significantly higher up to few orders of magnitude than any previous reported results.

## Acknowledgements:

The authors gratefully acknowledge the Science and Engineering Research Board (Project no: EMR/2016/002520) and DAE BRNS (Sanction no: 37(3)/14/26/2016-BRNS/37245) and FIST Project for Department of Physics. D. Mandal gratefully acknowledges CSIR for their financial support.

**Table 1.** Interband relaxation time obtained for selected wavelengths following nanosecond TA measurements.

| Wavelength (nm) | $\tau_2$ (µs) |
|:---:|:---:|
| 450 | 1.01± 0.01 |
| 500 | 1.38 ± 0.02 |
| 550 | 1.40 ± 0.07 |
| 600 | 1.92 ± 0.18 |

**Table 2.** Nonlinear parameters obtained from the fitting of the experimental data shown in Fig. 4(a)

| Intensity (MW/cm$^2$) | $\beta \times 10^6$ (cm/GW) | $I_S \times 10^{-4}$ (GW/cm$^2$) |
|---|---|---|
| 10 | $4.3 \pm 0.7$ | $37.7 \pm 0.5$ |
| 20 | $3.2 \pm 0.3$ | $38.5 \pm 0.3$ |
| 30 | $2.71 \pm 0.05$ | $45 \pm 0.2$ |

**Table 3.** Comparison of non-linear absorption coefficient (β) and refractive index ($n_2$) with previously reported samples when excited with 532 nm (2.33 eV) laser. OR and NR refer to off-resonant and near-resonant excitation, respectively.

| Sample | Intensity | Bandgap /excitation type | Pulse duration | β (cm/GW) | $n_2$ ($cm^2$/GW) | Ref |
|---|---|---|---|---|---|---|
| $Ga_5Sb_{10}Ge_{25}Se_{60}$ | 200 MW/$cm^2$ | 2.9 eV/OR | 7 ns | 400 | | 28 |
| $Ge_{28}Sb_{60}Se_{12}$ | 178 W/$m^2$ | 2.19 eV/NR | 7 ns | 17.92 | | 29 |
| ZnO | 360 MW/$cm^2$ | 3.01 eV/OR | 5 ns | 7.6 | | 30 |
| $\alpha$-$NiMoO_4$ | 210 MW/$cm^2$ | 3 eV/OR | 7ns | 71 | | 31 |
| Neutral red dye | 50 MW/$cm^2$ | 2.64 eV/NR | 30 ns | 1100 | $1.5 \times 10^{-3}$ | 32 |
| GO | 160 MW/$cm^2$ | 4.1 eV/OR | 5 ns | 20 | | 13 |
| $MoSe_2$/GO | 160 MW/$cm^2$ | broadband | 5 ns | 580 | | 13 |
| $As_{50}S_{50}$ | 10 MW/$cm^2$ | 2.05 eV /NR | 7 ns | $4.3 \times 10^6$ | | Present Work |

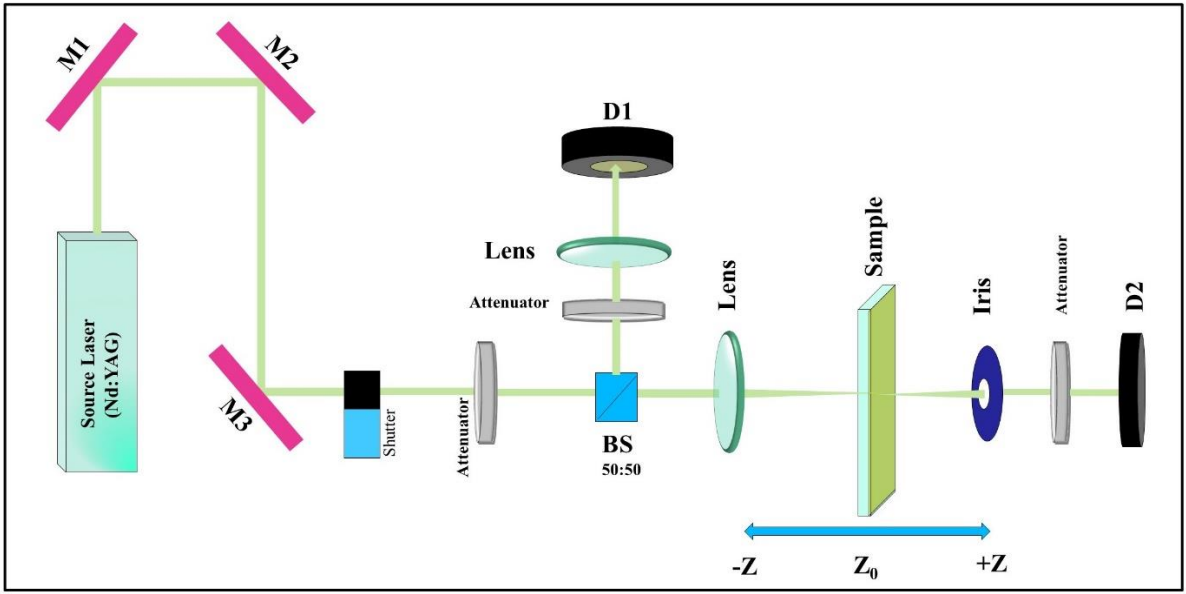

**Fig. 1.** Schematic diagram of Z-scan measurement.

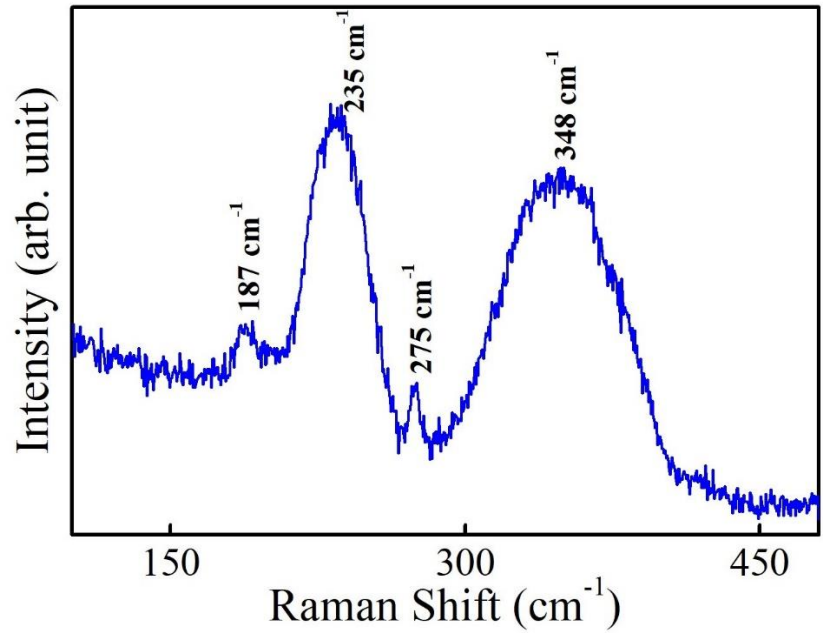

**Fig. 2.** Raman spectrum of $As_{50}S_{50}$ thin films.

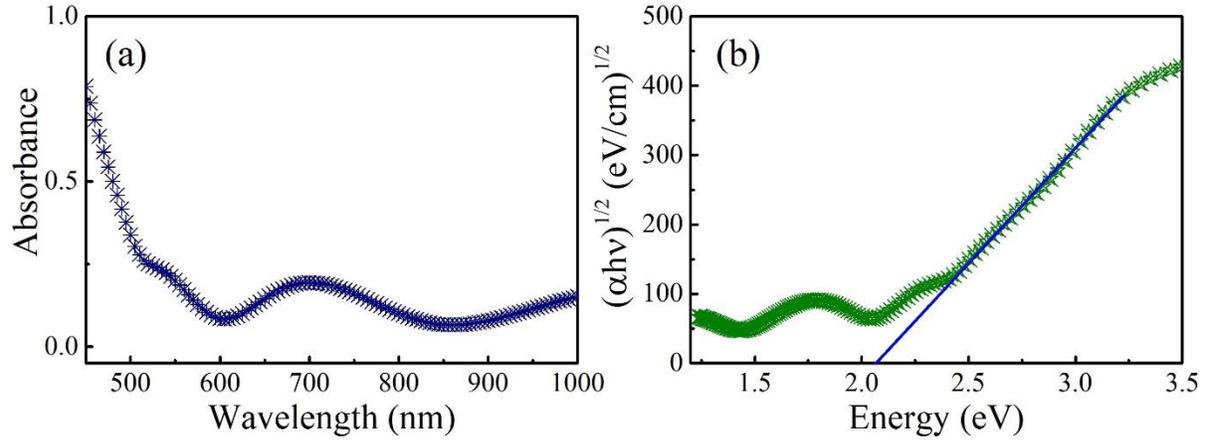

**Fig. 3.** (a) Optical absorption spectrum of $As_{50}S_{50}$ thin film. (b) Tauc plot used for calculating the band-gap.

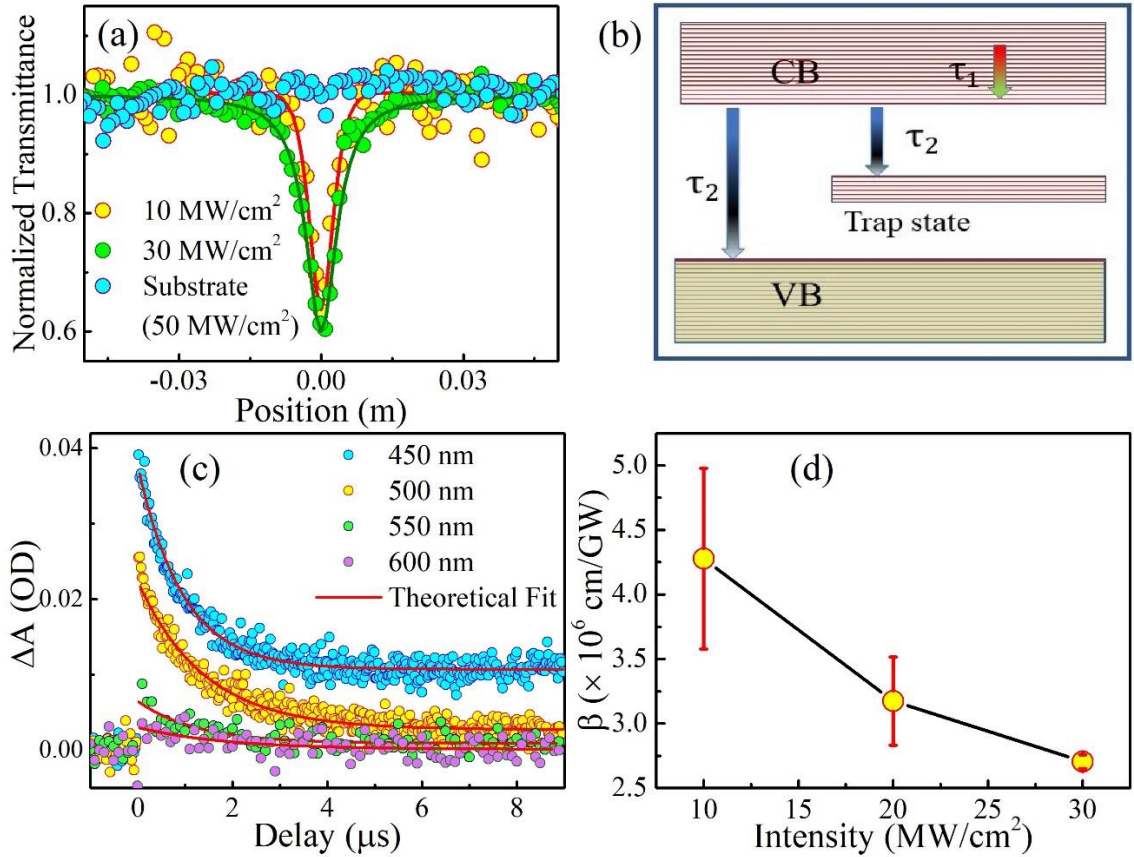

**Fig. 4.** (a) Intensity dependent open aperture Z-scan of As$_{50}$S$_{50}$ thin film, the horizontal blue line represents absence of nonlinear response from the quartz glass plate. (b) Energy level diagram for the excited state absorption. (c) Nanosecond transient absorption kinetics for selected wavelengths. (d) Variation of ESA coefficient as a function of input intensity.

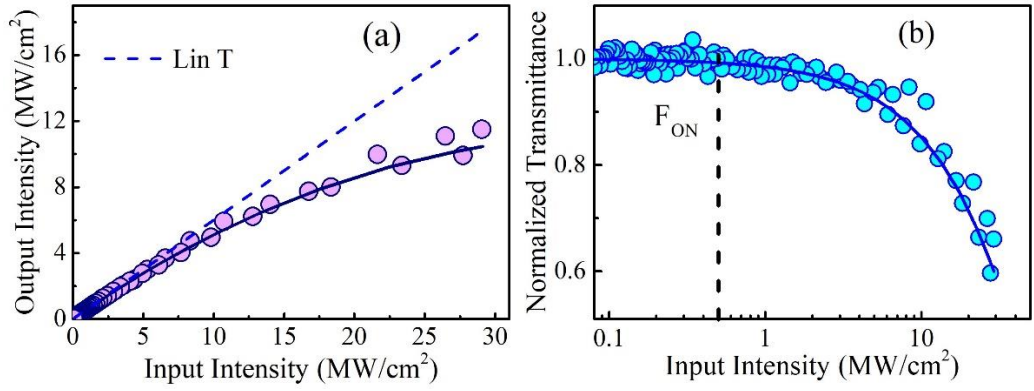

**Fig. 5.** (a) Output intensity as a function of input intensity. The output intensity deviates from the input intensity after a threshold of 5 MW/cm². (b) Normalized transmittance as a function of input intensity.

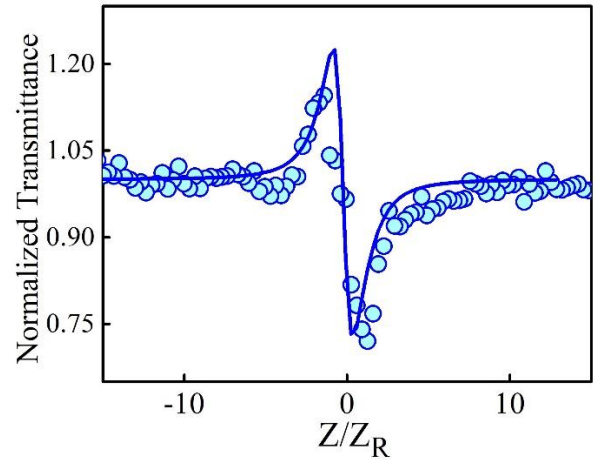

**Fig. 6.** Close aperture Z-scan traces of As$_{50}$S$_{50}$ thin films at 30 MW/cm$^2$. The hollow circles and solid line represent experimental data and theoretical fit, respectively.